\documentclass[sigconf]{acmart}
\usepackage{booktabs} % For formal tables
\usepackage{multirow}
\usepackage{subcaption}
\usepackage{caption}
\usepackage{setspace}
\usepackage{amsmath}
\usepackage[english]{babel}
\usepackage{float}
\usepackage{flushend}
\usepackage{mathtools}

\begin{document}

\copyrightyear{2019} 
\acmYear{2019} 
\acmConference[ICTIR '19]{The 2019 ACM SIGIR International Conference on the Theory of Information Retrieval}{October 2--5, 2019}{Santa Clara, CA, USA}
\acmBooktitle{The 2019 ACM SIGIR International Conference on the Theory of Information Retrieval (ICTIR '19), October 2--5, 2019, Santa Clara, CA, USA}
\acmPrice{15.00}
\acmDOI{10.1145/3341981.3344213}
\acmISBN{978-1-4503-6881-0/19/10}

\title{Neural Document Expansion with User Feedback}

\author{Yue Yin}
\authornote{This work was done when the first author was visiting Tsinghua University.}
\affiliation{
\institution{
Beijing Normal University}
}
\email{bnuyinyue@outlook.com}

\author{Chenyan Xiong}
\affiliation{
\institution{
Microsoft Research AI}
}
\email{chenyan.xiong@microsoft.com}

\author{Cheng Luo}
\affiliation{
\institution{
Tsinghua University}
}
\email{chengluo@tsinghua.edu.cn}

\author{Zhiyuan Liu}
\affiliation{
\institution{
Tsinghua University}
}
\email{liuzy@tsinghua.edu.cn}

\keywords{
Neural Information Retrieval;
Document Expansion;
User Feedback.
}

\begin{abstract}
This paper presents a neural document expansion approach (\texttt{NeuDEF}) that enriches document representations for neural ranking models. \texttt{NeuDEF} harvests expansion terms from queries which lead to clicks on the document and  weights these expansion terms with learned attention. It is plugged into a standard neural ranker and learned end-to-end. Experiments on a commercial search log demonstrate that \texttt{NeuDEF} significantly improves the accuracy of state-of-the-art neural rankers and expansion methods on queries with different frequencies. Further studies show the contribution of click queries and learned expansion weights, as well as the influence of document popularity of \texttt{NeuDEF}'s effectiveness.
\end{abstract}

\maketitle

\section{Introduction}

Neural information retrieval (Neu-IR) methods have shown promising results in various search scenarios.
These neural ranking models leverage distributed representations (embeddings) to conduct soft matches between query-documents, and at the same time, leverage large model capacity to revive ''classic'' IR intuitions,  such as translation model~\cite{K-NRM}, phrase matches~\cite{dai2018convolutional}, and multi-field evidence combination~\cite{zamani2018neural}.
This paper revisits the document expansion technique and develops \texttt{NeuDEF}, a Neural Document Expansion method that explicitly expands documents using user Feedback signals.
% for Neu-IR.

\texttt{NeuDEF} first harvests candidate expansion terms for a document from queries lead to user clicks on it (click queries). Then its \emph{attention mechanism} weights the expansion terms based on both self-attention and the matches between the document and click queries.
The weighted expansion terms form an additional document representation and can be integrated into various neural ranking models. During learning, the attention mechanism on expansion terms and the neural ranking model are end-to-end trained using document ranking labels; the integrated system learns how to expand and rank documents jointly.

In our experiments on two search log samples from a commercial search engine, \texttt{NeuDEF} significantly improves the ranking accuracy of its base ranker: \texttt{K-NRM}~\cite{K-NRM} and outperforms previous state-of-the-art neural ranking methods and document expansion techniques by large margins. 
Our additional studies show that \texttt{NeuDEF}'s attention mechanism assigns higher weights to novel expansion terms and \texttt{NeuDEF} generalizes user feedback signals to unseen queries.

\section{Neural Document Expansion}

This section presents \texttt{K-NRM}~\cite{K-NRM}, the base ranker, \texttt{NeuDEF}, our expansion model, and the joint learning of the two.

\subsection{Base Ranker Recap}
\texttt{K-NRM} is an interaction-based neural ranking model that uses density estimation kernels to soft match term pairs~\cite{K-NRM}:
\begin{align}
\text{K-NRM}(q, d; w) &= w^T \sum_{t_j \in q} \log \Phi(t_j, d), \\
\Phi(t_j, d) &= \{\phi_1(t_j, d),...\phi_k(t_j, d)..., \phi_K(t_j, d)\}, \\
\phi_k(t_j, d) &= \sum_{t_d \in d}\exp(-\frac{(\cos(\vec{t}_j, \vec{t}_d)-\mu_k)^2}{2 \sigma_k^2}). \label{eq:original_kernel}
\end{align}
$\phi_k$ is a Gaussian (RBF) kernel. It ''soft counts'' the number of document terms $t_d$ that match the query term $t_j$ near its kernel region $(\mu_k, \sigma_k)$. 
The match is calculated by the cosine similarity, $\cos(\vec{t}_j, \vec{t}_d)$, of their word embeddings, which are learned end-to-end for the whole vocabulary. 
$w$ is the kernel weight to be learned.

\subsection{Expansion Term Selection and Weighting}
\texttt{NeuDEF} leverages user feedback in search logs to find expansion terms.
It first considers the click queries $C_d$ of the document $d$: 
\begin{align}
C_d=  \{c_i | click(c_i, d) = \text{True}\}.
\end{align}
$click(c, d)$ is True if there is any click on $d$ from the query $c$.

Terms from clicked queries form the candidate  expansion set $T_d$:
\begin{align}
T_d = \{t_j | \exists c_i: t_j \in c_i, c_i \in C_d\}.
\end{align}
A clicked query is likely to be related to the document as a user has used the query to search and click on the document.

Given the expansion terms $T_d$ and clicked queries $C_d$ of document $d$, \texttt{NeuDEF}  calculates the weights $A(T_d, d)=\{a(t_j, d) | t_j \in T_d\}$ of the expansion terms $T_d$ by an attention mechanism. 

The attention first uses Multi-Head Self Attention~\cite{vaswani2017attention} to weight document and clicked queries independently. To capture the cross-queries information and generate better word-level attention weights, we concatenate all the clicked queries for a specific document and feed into the transformer (Self-ATT).
Then the attention matches the clicked queries to the document:
\begin{align}
m(c_i, d) &= \text{K-NRM}(\text{Self-ATT}(c_i), \text{Self-ATT}(d); w_c). \label{eq:att_knrm}
\end{align}
It uses the same architecture as the base ranker but different parameters.
Then it calculates the attention weight of term $t_j$ by summing match scores from the clicked queries it appears in:
\begin{align}
a(t_j, d) &= \sum_{c_i, t_j \in c_i} m(c_i, d). \label{eq:att}
\end{align}
The more related clicked queries the term $t_j$ appears in, the more expansion weights, $a(t_j, d)$, it receives.

\texttt{NeuDEF} provides the expansion terms $T_d$, which introduce user feedback, and their weights $A(T_d, d)$ learned by self-attention and document-query matches.

\subsection{Joint Learning with Neural Rankers}

\texttt{NeuDEF} is integrated to the base ranker $f$ by providing an additional expansion field, which includes expansion terms $T_d$ and attention weights $A(T_d, d)$ and is linearly combined with the base ranker ($\alpha$ and $\beta$ are the combine weights):
\begin{align}
f_{\texttt{NeuDEF}}(q, d) &= \alpha f(q, d) + \beta f'(q, de), \label{eq:rank}\\
f(q, d) &= \text{K-NRM}(q, d; w_r).
\end{align}

We use a modified K-NRM on the expansion field: 
\begin{align}
f'(q, de) &= w_{de}^T \sum_{t_q \in q} \log \Phi'(t_q, de), \label{eq:f_qde} \\
\phi_k'(t_q, de) &= \sum_{t_j \in T_d} a(t_j, d) \exp(-\frac{(\cos(\vec{t}_q, P\vec{t}_j) - \mu_k)^2}{2\sigma_k^2}). \label{eq:kernel}
\end{align}
The modified version weights expansion term $t_j$ by the attention $a(t_j, d)$, and adds a projection $P$ on the expansion word embeddings to distinguish them from the original words. The three \texttt{K-NRM} models used in $f(q, d)$, $f'(q, de)$, and $m(c_i, d)$ (Eq.~\ref{eq:att_knrm}) share the same word embeddings and kernel hyper-parameters.

\texttt{NeuDEF} is then trained with the neural ranker using standard pairwise hinge loss:
\begin{align}
\sum_{d^+, d^- \in D^{+,-}} 
\max(0, 1 - f_{\texttt{NeuDEF}}(q, d^+) + f_{\texttt{NeuDEF}}(q, d^-)) \label{eq:loss}.
\end{align}
$d^+, d^-$ are the relevant and irrelevant document pairs of the query. 
Instead of conducting the expansion and ranking separately,
\texttt{NeuDEF} learns the document expansion and document ranking jointly from ranking labels using back-propagation.

\section{Experimental Methodology}

\textbf{Dataset.} Sogou, a commercial search engine based in China, released a sample of search log with queries, documents, and user clicks to various academic partners.
Our experiments use two training datasets sampled from Sogou log: 
\texttt{Sogou-KNRM}, the one used by \texttt{K-NRM}~\cite{K-NRM}, for a fair comparison, and \texttt{Sogou-QCL}, the public release of Sogou search log~\cite{zheng2018sogou}. We follow the same setting with prior research~\cite{K-NRM, zheng2018sogou} and
refer to their papers for more details in the datasets due to space limitation~\cite{K-NRM, zheng2018sogou}.

Our evaluations on \texttt{Sogou-KNRM} dataset omit the Testing-SAME setting which is prune to overfitting~\cite{dai2018convolutional} and add torso (50-1000 appearances) and tail (less than 50) queries. We use four evaluation scenarios for \texttt{Sogou-KNRM} dataset: (1) \textbf{Testing-Raw} \textbf{Head}, \textbf{Torso}, and \textbf{Tail}: User clicks as relevance labels and evaluate on head, torso, and tail queries; (2) \textbf{Testing-DIFF}: Click model (TACM~\cite{liu2017time})  inferred relevance labels and evaluate on head queries.
For \texttt{Sogou-QCL} dataset, we use TACM inferred relevance labels to train and test our model on head queries, following the exact same setting for the Table 5 in \texttt{Sogou-QCL} original paper~\cite{zheng2018sogou}. 

To study the effectiveness of document expansion with body field,
The body texts are from our crawled HTMLs and parsed by Boilerpipe. They are combined linearly with the titles following standard multi-field ranking setup.

\begin{table*}
\caption{Ranking Accuracy on Sogou-KNRM dataset.
Relative performances over \texttt{K-NRM} are in percentages.
$\dagger$, $\ddagger$, $\mathsection$ and $^\mathparagraph$ indicate statistically significant improvements over 
\texttt{K-NRM}$^\dagger$, 
\texttt{NRM-F}$^\ddagger$,
\texttt{NeuDEF-TF}$^\mathsection$,
and \texttt{NeuDEF-NoTrans}$^\mathparagraph$.
% Both document title and body text are used.
\label{tab:overall}
}
\centering
\begin{tabular}{ l |l r| l r| l r| l r |l r}
\hline
\hline
\bf{}  &\multicolumn{6}{|c}{\bf{Testing-RAW, MRR}} & \multicolumn{4}{|c}{\bf{Testing-DIFF}}  \\ \hline
\bf{Model}  & \multicolumn{2}{|c}{\bf{Head}} & \multicolumn{2}{|c}{\bf{Torso}} & \multicolumn{2}{|c}{\bf{Tail}} & \multicolumn{2}{|c}{\bf{NDCG@1}} & \multicolumn{2}{|c}{\bf{NDCG@10}} \\ 
\hline
%\texttt{BM25}~\cite{robertson1999okapi}     &  0.2280 & -31.7\% & 0.2766 & -25.1\%  & 0.2640 &  -16.2\%& 0.1631& -44.4\% &  0.3254& -22.3\%     \\ 
%\texttt{CoorAscent}~\cite{metzler2007linear} &  0.2415 & -27.7\% & 0.3202 & -13.2\% &0.2977 & -5.6\%  & 0.2089 & -28.7\%       & 0.3775& -9.9\% \\ 
%\hline
%\texttt{MP}~\cite{pang2016text}    & 0.2404 & -28.0\% & 0.3094 & -16.2\%  &  0.2958& -6.2\%   & 0.1969 & -32.8\%   & 0.3450 & -17.7\%  \\
\texttt{DRMM }~\cite{jiafeng2016deep}    & 0.2335 & -30.1\% & 0.3102 & -16.0\%  &  0.2951&  -6.4\%   & 0.2126 & -27.5\%   & 0.3592 & -14.3\% \\
\texttt{CDSSM}~\cite{shen2014learning}    & 0.2501 & -25.1\% & 0.3184 & -13.7\%  & 0.2928 & -7.1\%   & 0.2017  & -31.2\%    & 0.3500  &  -16.5\%  \\
\hline
\texttt{K-NRM}~\cite{K-NRM}      & 0.3339 & - & 0.3691 &- &  0.3152 & -& 0.2931 & - & 0.4190 &-   \\ 
\texttt{Conv-KNRM}~\cite{dai2018convolutional}      & 0.3382 & +1.3\% & 0.3645 & -1.2\% & 0.3218 & +2.1\%& 0.2988 & +1.9\% & 0.4204 &+0.3\%   \\ 
\hline
\texttt{K-NRM+DELM+TF}~\cite{tao2006language}      & 0.3351  & +0.4\% & 0.3701 & +0.3\% &  0.3121 & -1.0\% &  0.2901& -1.0\%  & 0.4203 &   +0.3\%  \\ 
\texttt{K-NRM+ExpaNet}~\cite{tang2017end}      &  0.3402 & +1.9\% & 0.3702 & +0.3\% & 0.3234  & +2.6\% & 0.3004 & +2.5\%  & 0.4212 &   +0.5\%  \\ 
\hline
% \texttt{K-NRM+URL}      & $0.3599^{\dagger}$ & +7.8\% & 0.3703 & +0.3\% &  $0.3302^{\dagger}$ & +4.8\% &  $0.3259^{\dagger}$ & +11.2\%  & $0.4404^{\dagger}$ &   +5.1\%  \\ 
\texttt{K-NRM+DocFreq}      &  $0.3501^{\dagger}$ & +4.9\% & 0.3714 & +0.6\% &  $0.3297^{\dagger}$ & +4.6\% & $0.3223^{\dagger}$ & +10.0\%  &  0.4289&   +2.4\%  \\ 
\texttt{K-NRM+CQCount}      & $0.3604^{\dagger}$  & +7.9\% & 0.3785 & +2.5\% & $0.3386^{\dagger}$  & +7.4\% &  $0.3345^{\dagger}$ & +14.1\%  &  $0.4398^{\dagger}$ & +5.0\%  \\ 
% \texttt{K-NRM+DocID}      & $0.3623^{\dagger}$  & +8.5\% & $0.3821^{\dagger}$ & +3.5\% & $0.3397^{\dagger}$  & +7.8\% & $0.3302^{\dagger}$ & +12.7\%  &  $0.4501^{\dagger}$ & +7.4\%  \\ 
\hline
\texttt{NRM-F}~\cite{zamani2018neural}    & $0.3747^{\dagger}$ & +12.2\% & $0.4094^{\dagger}$ & +10.9\%  & $0.3545^{\dagger}$ & +12.5\%   &  $0.3419^{\dagger}$ & +16.6\%    &  $0.4776^{\dagger}$ &  +14.0\%  \\
\hline
\texttt{NeuDEF-TF}    &  $0.3947^{\dagger, \ddagger}$ & +18.2\% & $0.4246^{\dagger}$ & +15.0\%  & $0.3430^{\dagger}$  & +8.8\%   &  $0.3672^{\dagger, \ddagger}$ & +25.3\%    & $0.4896^{\dagger, \ddagger}$  &  +16.8\%  \\
\texttt{NeuDEF-NoTrans}    &  $\bf{0.4054}^{\dagger, \ddagger, \mathsection}$ & +21.4\% &  $0.4688^{\dagger, \ddagger, \mathsection}$ & +27.0\%  & $0.3584^{\dagger}$ & +13.7\%   &  $0.3785^{\dagger, \ddagger}$ & +29.1\%    & $0.5023^{\dagger, \ddagger, \mathsection}$  &  +19.9\%  \\
\texttt{NeuDEF}  & $0.4038^{\dagger, \ddagger, \mathsection}$ & +20.9\% & $\bf{0.4730}^{\dagger,\ddagger,\mathsection}$ & +28.1\%  & $\bf{0.3675}^{\dagger,\ddagger, \mathsection, \mathparagraph}$ & +16.6\%   & $\bf{0.3858}^{\dagger, \ddagger}$  & +31.6\% & $\bf{0.5056}^{\dagger, \ddagger, \mathsection}$  &  +20.7\%  \\
\hline
\hline
\end{tabular}  
\end{table*}

\begin{table*}
\caption{Performance on title (T) and body (B) field individually on Sogou-KNRM. Relative performances compared to \texttt{K-NRM} and the significant improvements over \texttt{K-NRM}$^\dagger$, \texttt{NRM-F}$^\ddagger$, \texttt{NeuDEF-TF}$^\mathsection$ and \texttt{NeuDEF-NoTrans}$^\mathparagraph$ are compared in each field group.
\label{tab:overall_content}
}
\centering
\begin{tabular}{ l |l r| l r| l r| l r |l r}
\hline
\hline
\bf{}  &\multicolumn{6}{|c}{\bf{Testing-RAW, MRR}} & \multicolumn{4}{|c}{\bf{Testing-DIFF}}  \\ \hline
\bf{Model}  & \multicolumn{2}{|c}{\bf{Head}} & \multicolumn{2}{|c}{\bf{Torso}} & \multicolumn{2}{|c}{\bf{Tail}} & \multicolumn{2}{|c}{\bf{NDCG@1}} & \multicolumn{2}{|c}{\bf{NDCG@10}} \\ 
\hline

\texttt{K-NRM(T)}  & 0.3440 & -   & 0.3747 & -      & 0.3244 & -      & 0.3132 & -       & 0.4288 & - \\ 
\texttt{NRM-F(T)}  & $0.3835^{\dagger}$ & +11.5\% & $0.4174^{\dagger}$ & +11.4\% & $0.3448$ & +6.3\% & $0.3706^{\dagger}$ & +18.3\% & $0.4651^{\dagger}$ & +8.5\%  \\
\texttt{NeuDEF-TF(T)} & $0.3905^{\dagger}$ & +13.5\% & $0.4523^{\dagger, \ddagger}$ & +20.7\% & $0.3322$ & +2.4\% & $0.3678^{\dagger}$ & +17.4\% & $0.4864^{\dagger}$ & +13.4\%  \\
\texttt{NeuDEF-NoTrans(T)} & $\bf{0.4104}^{\dagger,\ddagger,\mathsection}$ & +19.3\% & $0.4692^{\dagger,\ddagger}$ & +25.2\%  & $0.3608^{\dagger,\mathsection}$ & +11.2\% & $0.3759^{\dagger}$  & +20.0\% &  $0.4985^{\dagger,\ddagger}$ &   +16.3\%  \\
\texttt{NeuDEF(T)} & $0.4017^{\dagger,\ddagger,\mathsection}$ & +16.8\% & $\bf{0.4775}^{\dagger,\ddagger}$ & +27.4\%  & $\bf{0.3706}^{\dagger,\ddagger, \mathsection, \mathparagraph}$ & +14.2\% & $\bf{0.3763}^{\dagger}$  & +20.1\% &  $\bf{0.5003}^{\dagger,\ddagger}$ &   +16.7\%  \\
\hline
\texttt{K-NRM(B)}   & 0.2728 & - & 0.3226 & - & 0.2539 & - & 0.2275 & - & 0.3744 &-   \\ 
\texttt{NRM-F(B)}   & $0.3486^{\dagger}$ & +27.8\% & $0.3907^{\dagger}$ & +21.1\%  & $0.3147^{\dagger}$ & +23.9\%   & $0.3092^{\dagger}$ & +35.9\%  & $0.4601^{\dagger}$ &  +22.9\%  \\
\texttt{NeuDEF-TF(B)} & $0.3608^{\dagger}$ & +32.3\% & $0.3945^{\dagger}$ & +22.3\%  & $0.3164^{\dagger}$ & +24.6\%   & $0.3335^{\dagger}$ & +46.6\%  & $\bf{0.4730}^{\dagger}$ &  +26.3\%  \\
\texttt{NeuDEF-NoTrans(B)}  & $0.3605^{\dagger}$ & +32.1\% & $0.4140^{\dagger,\ddagger,\mathsection}$ & +28.3\%  & $0.3279^{\dagger}$ & +29.1\%   & $0.3304^{\dagger}$  & +45.2\% & $0.4729^{\dagger}$  &  +26.3\%  \\
\texttt{NeuDEF(B)}  & $\bf{0.3633}^{\dagger}$ & +33.2\% & $\bf{0.4309}^{\dagger,\ddagger,\mathsection,\mathparagraph}$ & +33.6\%  & $\bf{0.3397}^{\dagger,\ddagger,\mathsection,\mathparagraph}$ & +33.8\%   & $\bf{0.3338}^{\dagger}$  & +46.7\% & $0.4715^{\dagger}$  &  +25.9\%  \\
% \hline
% \texttt{K-NRM(T+B)}  & 0.3339 & - & 0.3691 & - & 0.3152 & - & 0.2931 & - & 0.4190 & -   \\ 
% \texttt{NRM-F(T+B)}  & $0.3747^{\dagger}$ & +12.2\% &  $0.4094^{\dagger}$ & +14.6\%  &  $0.3545^{\dagger}$ & +12.5\%   & $0.3419^{\dagger}$ & +16.6\% &  $0.4776^{\dagger}$ &  +14.0\%  \\
% \texttt{NeuDEF(T+B)} & $0.4054^{\dagger,\ddagger}$ & +21.4\% &  $0.4688^{\dagger,\ddagger}$ & +27.0\%  & $0.3584^{\dagger}$  & +13.7\%   &  $0.3785^{\dagger,\ddagger}$ & +29.1\%  & $0.5023^{\dagger,\ddagger}$  &  +19.9\%  \\
\hline
\hline
\end{tabular}  
\end{table*}

\textbf{Expansion Candidates.} All expansion approaches harvest the expansion terms \emph{solely using the queries and clicks in the training split}. As there is no overlap in the training and testing queries, the expansion approaches use no information from the testing data.

\textbf{Evaluation Metrics.}
Testing-DIFF is evaluated by NDCG@\{1, 10\} and Testing-Raw is evaluated by MRR~\cite{K-NRM}.
Statistical significance is tested by permutation test with $p<0.05$. 

Model performances of the ranking model ($f_2$) w.r.t the baseline model ($f_1$) at \emph{per document level} are compared by $\Delta$RR:
\begin{align}
\Delta \text{RR}_{f_1 \rightarrow f_2}(d) &= \sum_{q} y(q,d) \{RR_{f_2}(q, d) - RR_{f_1}(q, d)\}.
\end{align}
$y(q,d)=\{+1, -1\}$ is the relevance label. RR(q, d) is the reciprocal rank of d under q.
Better ranking models have positive $\Delta$RRs.

\textbf{Baselines.} 
Using the same experimental setup with prior research~\cite{K-NRM, zheng2018sogou} makes our method directly comparable with their baselines: \texttt{CDSSM}~\cite{shen2014learning},
\texttt{DRMM}~\cite{jiafeng2016deep}, \texttt{K-NRM}~\cite{K-NRM} and \texttt{Conv-KNRM}~\cite{dai2018convolutional}. We use their shared implementations to obtain baseline results on torso and tail queries.
We also implemented \texttt{NRM-F}~\cite{zamani2018neural}, the fielded version of \texttt{CDSSM}, using the same fields as in \texttt{NeuDEF}.
All neural ranking baselines leverage user feedback following Eq.~\ref{eq:rank}.

We implemented and compared with many document expansion baselines.
\texttt{DELM}~\cite{tao2006language} is a traditional document expansion language model via document neighbors. We selected the top 5 words as document expansion fields according to their frequency in all the neighbor documents. 
\texttt{ExpaNet}~\cite{tang2017end} is a neural text expansion model with memory network generated via document neighbors. 
We combined the expanded document features from \texttt{ExpaNet} with soft match features in original \texttt{K-NRM}'s dense layer. We treated all the documents under a specific query as neighborhoods. 

We also compared with expanding using other meta-data:
(1) DocFreq: the number of times the document appears in search log;
(2) CQCount: the number of queries lead to clicks in the document;

Two simpler versions of \texttt{NeuDEF} are compared too: (1) \texttt{NeuDEF-TF}, which weights expansion terms only by their frequency in the clicked queries; (2) \texttt{NeuDEF-NoTrans}, which does not use transformer's self attention.

\textbf{Implementation Details.}
All baselines use the same setting in prior research~\cite{K-NRM}: 300-dimension embedding layer; 165,877-word vocabulary; one exact match kernel ($\mu=1, \sigma=10^{-3}$); and ten kernels equally distributed in (-1, 1) ($\mu \in \{0.9, 0.7,...,-0.9\}, \sigma=0.1$).
1 multi-head attention layer with 4 heads is used in \texttt{NeuDEF}. 
Documents with no clicked query are not expanded.
The learning of all methods use the same training data. 
All neural methods use Adam optimizer with learning rate $0.001$, batch size 64 and $\epsilon=1e-5$, and early stopping on $20\%$ random selected validation data.

\section{Evaluation}
This section presents evaluation results.

\subsection{Overall Ranking Accuracy}
Table~\ref{tab:overall} lists the results on \texttt{Sogou-KNRM}. 
\texttt{NeuDEF} outperforms all baselines.
It improves \texttt{NRM-F}, the previous state-of-the-art, on Head, Torso, and Tail.
 \texttt{NeuDEF} outperforms it base ranker (\texttt{K-NRM}) on tail queries by ($13.7\%$). 
The attention mechanism learns effective expansion weights: \texttt{NeuDEF} significantly outperforms \texttt{NeuDEF-TF}. The transformer helps on the tail, as compared with  \texttt{NeuDEF-NoTrans}.

Table~\ref{tab:overall_content} lists the results of \texttt{K-NRM}, \texttt{NRM-F}, and \texttt{NeuDEF}s on the title and body individually. \texttt{NeuDEF} performs the best on each field. Note that it has been observed that adding body fields does not contribute much~\cite{K-NRM, zamani2018neural}. How to better model long body text is a future research direction.
The performances of \texttt{NeuDEF} and main baselines on \texttt{Sogou-QCL}~\cite{zheng2018sogou} are in Table~\ref{tab:qcl_overall}. The trends are similar.

\subsection{Learned Expansion Weights}
\label{sec:exp_ablation}
\begin{figure*}
\centering
\begin{tabular}{ c c c }
\includegraphics[width=1.9in]{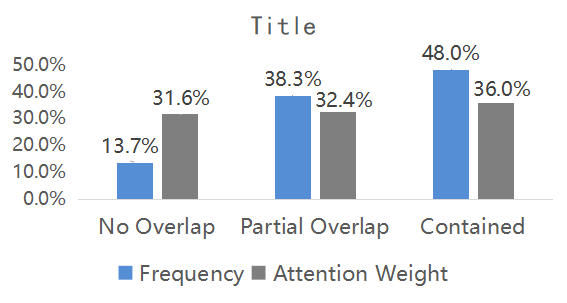} &   \includegraphics[width=1.9in]{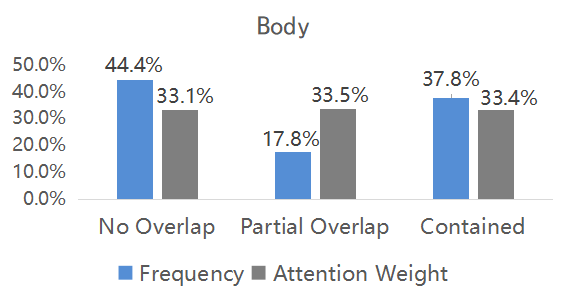} &
\includegraphics[width=1.9in]{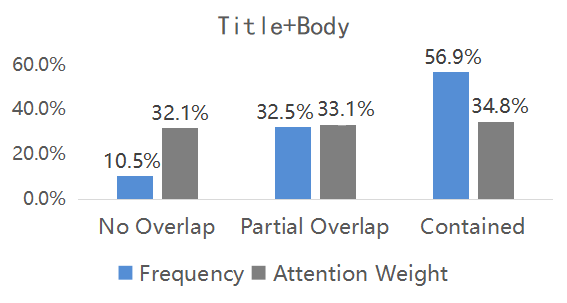} \\
\end{tabular}  
\caption{Frequency Distribution and normalized Attention Weights on the expansion terms from three clicked query groups: those have No Overlap with, Partial Overlap with, or are Contained by the corresponding document fields.  \label{fig:learn_exp} }
\end{figure*}

This experiment analyzes the expansion weights learned by \texttt{NeuDEF}'s attention mechanism on three groups of expansion terms: those from clicked queries that have  \emph{No Overlap} with the document content, those have \emph{Partial Overlaps}, and those \emph{Contained} by the document. 
Figure~\ref{fig:learn_exp} shows the distribution of terms in these three groups and their average expansion weights normalized on each document.

\begin{table}
\caption{Accuracy on Sogou-QCL dataset.
Relative performances are compared to \texttt{K-NRM}.
Statistical significance is marked by
$\dagger$(\texttt{K-NRM}), $\ddagger$(\texttt{NRM-F}), $\mathsection$(\texttt{NeuDEF-TF}) and $\mathparagraph$(\texttt{NeuDEF-NoTrans}).
\label{tab:qcl_overall}
}
\centering
\begin{tabular}{ l | l r |l r}
\hline
\hline
\bf{} & \multicolumn{4}{|c}{\bf{Testing-TACM}}  \\ \hline
\bf{Model}  & \multicolumn{2}{|c}{\bf{NDCG@1}} & \multicolumn{2}{|c}{\bf{NDCG@10}} \\ 
\hline
%\texttt{BM25}~\cite{robertson1999okapi} &  0.2018 & -16.2\% & 0.3682 & -5.3\%     \\ 
%\texttt{DRMM}~\cite{jiafeng2016deep}     & 0.1647  & -31.6\% & 0.3382 & -13.0\%     \\ 
\hline
\texttt{K-NRM}~\cite{K-NRM}     & 0.2409  & - & 0.3888 & -     \\ 
%\texttt{Conv-KNRM}~\cite{dai2018convolutional}      &  0.2497 & +3.7\% & $0.4012^{\dagger}$ & +3.2\%     \\ 
%\hline
\texttt{NRM-F}~\cite{zamani2018neural}      &  $0.2546^{\dagger}$ & +5.7\% &$0.4438^{\dagger}$  & +14.1\%     \\ 
\texttt{NeuDEF-TF}     & $0.2734^{\dagger, \ddagger}$  & +13.5\% &  $0.4606^{\dagger, \ddagger}$& +18.5\%     \\ 
\texttt{NeuDEF-NoTrans}     & $0.2801^{\dagger, \ddagger, \mathsection}$  & +16.3\% & $\bf{0.4651}^{\dagger, \ddagger}$ & +19.6\%     \\ 
\texttt{NeuDEF}     & $\bf{0.2804}^{\dagger, \ddagger, \mathsection}$  & +16.4\% & $0.4643^{\dagger, \ddagger}$ & +19.4\%     \\ 
\hline
\hline
\end{tabular}  
\end{table}

About $10\%$ clicked queries have no term overlap with the document title or body (they might be retrieved by some query expansion-alike techniques).
\texttt{NeuDEF} assigns about one thirds of its learned attention weights on clicked queries that have no overlap with the document; with document title, $10\%$ of expansion terms from the No Overlap group received $30\%$ of expansion weights. \texttt{NeuDEF} learns to favor novel expansion terms that are related to but do not appear in the document, which may bring in extra information.

\begin{figure}[t]
\centering
\includegraphics[width=0.32\textwidth]{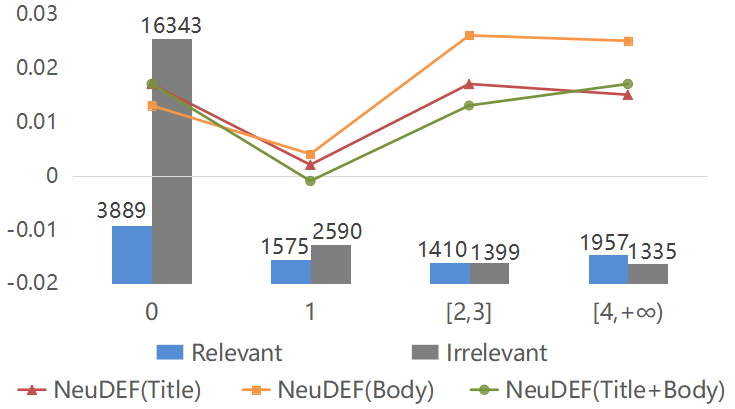} 
\caption{Performance on documents with different number of clicked queries.
X-axis is the number of clicked queries. Histograms are the number of documents.
Plots and Y-axis are the average $\Delta$RR compared to \texttt{K-NRM}; higher is better.
\label{fig:per_doc}
}
\end{figure}

\subsection{Document Level Performances}
\label{sec:doc_level}

In our experiments, all testing queries are ``unseen'' queries as they never appear in the training split nor used in the document expansion. 
The advantage of \texttt{NeuDEF} is that it leverages the feedback signals at document level: a query might never appear before in the query log but the candidate documents may have seen before.

This experiment studies \texttt{NeuDEF}'s document level performance w.r.t. different amounts of user feedback signals.
It groups documents based on their number of clicked queries, then it evaluates the $\Delta$RR of \texttt{NeuDEF} over \texttt{K-NRM} on each combination of document fields. 
The results on head queries are shown in Figure~\ref{fig:per_doc}.
Results on torso and tail are similar and omitted due to space constraints.

The number of clicked queries per document follows a long tail distribution. 
The user preferences also heavily favor popular documents; documents with more click queries are more likely to be relevant.
\texttt{NeuDEF} performs better than \texttt{K-NRM} on all groups and with all types of document content. 
Even on documents with no clicked queries where \texttt{NeuDEF} withdraws to the base \texttt{K-NRM} model, adding expansion terms provides extra information in training and helps \texttt{NeuDEF} learn better parameters than its base ranker.

\section{Conclusions and Future Work}

This paper presents \texttt{NeuDEF}, a neural document expansion approach that enriches document representations for neural ranking models using user feedback signals.
Experiments  demonstrate \texttt{NeuDEF}'s effectiveness and its ability to better utilize user feedback signals and generalize them to unseen queries through document expansions.
Future work includes bringing expansion terms from external resources and developing more advanced neural expansion models.

\section{Acknowledgement}

This work is supported by the National Key Research and Development Program of China (No. 2018YFC0831901).
We thank Sogou for providing access to the search log.

\bibliographystyle{ACM-Reference-Format}
\normalsize
\bibliography{main}
\flushend 

\end{document}